# Modified Apriori Approach for Evade Network Intrusion Detection System

Laxmi Lahoti
Department of Information Technology
MIT, Pune, Maharashtra, India
laxmil144@gmail.com

Chaitali Chandankhede
Department of Information Technology
MIT, Pune, Maharashtra, India
chaitali.chankhede@mitpune.edu.in

Debajyoti Mukhopadhyay
Department of Information Technology
MIT, Pune, Maharashtra, India
debajyoti.mukhopadhyay@gmail.com

*Abstract*— **Intrusion Detection System (IDS) is a software or hardware tool that repeatedly scans and monitors events that took place in a computer or a network. A set of rules are used by Signature based Network Intrusion Detection Systems (NIDS) to detect hostile traffic in network segments or packets, which are so important in detecting malicious and anomalous behavior over the network like known attacks that hackers look for new techniques to go unseen. Sometime, a single failure at any layer will cause the NIDS to miss that attack. To overcome this problem, a technique is used that will trigger a failure in that layer. Such technique is known as Evasive technique. An Evasion can be defined as any technique that modifies a visible attack into any other form in order to stay away from being detect. The proposed system is used for detecting attacks which are going on the network and also gives actual categorization of attacks. The proposed system has advantage of getting low false alarm rate and high detection rate. So that leads into decrease in complexity and overhead on the system. The paper presents the Evasion technique for customized apriori algorithm. The paper aims to make a new functional structure to evade NIDS. This framework can be used to audit NIDS. This framework shows that a proof of concept showing how to evade a self-built NIDS considering two publicly available datasets.**

## I. INTRODUCTION

Various sites install an Intrusion Detection System (IDS) to observe their hosts and networks for doubtful actions which are going. Many IDSs use a database of known events for comparison sending alarm when an equivalent action/event is detected. Nowadays, many organizations and companies use Internet services as their communication and marketplace to do business such as at EBay and Amazon.com website. Together with the increase of computer network activities the rising rate of network attacks has been advancing, impacting to the accessibility, confidentiality and integrity of important information data. Therefore a network system exploits one or more security tools such as firewall, antivirus, IDS and Honey Pot to avoid important data from illegal enterprise.

A network system using a firewall only is not enough to prevent networks from each and every attack types. The firewall cannot protect the network against intrusion attempts during the opening port. Hence an Intrusion Detection System is a prevention tool that gives an alarm signal to the computer user or network administrator for hostile activity on the opening session by inspects unsafe network activities.

Intrusion detection is a set of techniques and methods that are used to detect disbelieving activity both at the network and host level. A network based IDS (NIDS) processes any clear-text traffic that crosses the monitored network without degrading performance on the host computers, since a single NIDS can monitor several hosts, fewer maintenance and monitoring attempt is required. Network-based IDS cannot precisely know the target's machine state; it must instead deduce the effects of traffic on the target system. In contrast, a host-based IDS (HIDS) is installed on individual hosts, which grants knowledge of the target machine's state and the ability to sense attacks from any point of entrance. Network-based intrusion detection systems continue to be more prevalent and mature than their host-based counterparts, although personal firewalls such as Zone Alarm on Windows computers have host-based intrusion detection capability and are frequently in use. NIDS will only report whether a known attack was launched without being able to determine whether the attack succeeded or indeed whether the attack even applied to the target's operating system.

A common frustration for NIDS operators is the high level of false positives triggered on busy networks. Sometimes the NIDS reports an attack that has no applicability to the target, or that has simply unsuccessful, while other times the NIDS confuses innocuous traffic with misuse. Operators learn that they must tune their NIDS to reduce the number of false positive alerts on their particular network. Sites differ NIDS sensitivity in different areas, dipping the sensitivity to decline false positives.

## II. OVERVIEW

In this section we provide an overview of the problem and our result. We offer motivation of our work through a realistic scenario.

*A. Existing System*

As stated earlier Host based approaches detect intrusions utilizing audit data that are collected from the target host machine. As the information given by the review data can be tremendously inclusive and complicated. Hybrid systems as the name suggests is a system that is developed usually using a combination of both for e.g. host-based and network-based systems.

In a previous work [1], GP was used to model a simple NIDS with high false alarm rate. In this paper we present new improvements, performing evasions over that NIDS and corroborating the effectiveness of modeling NIDS with apriori approach.

Evasions on NIDS were first proposed by Ptacek and Newsham in 1998 [2]. In this seminal paper, the authors highlighted the existence of some ambiguities in the TCP and IP protocols, which allow different systems to implement them in a different way. An evasion succeeds when NIDS ignore packets which are going to be processed on the endpoints or vice versa. Many techniques have been designed to prevent evasions. Most of them are based on network traffic modification, to remove the ambiguities and establish a common understanding of the protocols for NIDS and endpoints. Watson et. al [7] propose a system called Protocol Scrubbing that generates well formed TCP data from traffic. A similar approach was proposed by Handley et al. [7], who introduced the concept of traffic normalizes. Those are intermediate elements that are placed in networks segments to remove possible ambiguities before being exposed to the NIDS. Because some of the evasive techniques are based on packet fragmentation and reassembly, the state of each connection and the previous packets must be stored and processed, in order to analyze the consistency of the connection. This situation consumes a large quantity of resources, leading into a bottleneck when working with high speed networks [9]. Some other solutions that do not modify the traffic have also been proposed. Varguese et al. [10] present the idea of dividing the entire signature of the NIDS into single smaller strings. A fast path finds matches with them and a slower one inspects it deeper if any match is found. Shankar and Paxon [11] propose a system that reports the NIDS about network topologies and the interpretation policy of the endpoint being monitored. Thus, NIDS can adapt their configuration taking into account that information. Snort [10] has adopted this technique in its IP processor.

The disadvantages of existing system are as follows:

Host-based IDSs are harder to organize, as information must be configured and managed for every host monitor.
ii. Since at least the information sources (and sometimes part of the analysis engines) for host-based IDSs be real on the host targeted by attacks, the IDS may be attack and disable as portion of the attack.
iii. Host-based IDSs are not well matched for detecting network scans or other such observation that targets an entire network because the IDS only sees those network packets received by its host.
iv. Host-based IDSs can be disabled by certain denial-of-service attacks.

*B. Need of Innovation*

Intrusion detection allows organizations to protect their systems from the threats that come with growing network connectivity and dependence on information systems. Given the point and environment of modern network security threats, the difficulty for security professionals should not be whether to use intrusion detection but which intrusion detection features and means to use. IDSs have gained receipt as a necessary addition to every organization's security infrastructure. In spite of the documented contributions intrusion detection technologies make to system security, in many organizations one must still defend the gaining of IDSs.

There are several forceful reasons to acquire and use IDS:
1. To prevent problem behaviors by raising the seeming risk of discovery and retribution for those who would attack or otherwise neglect the system
2. To distinguish attacks and other security violations that is not disallowed by other security measures
3. To identify and deal with the preamble to attack
4. To document the accessible threat to an organization.
5. To proceed as quality control for defense mean and administration, especially of huge and complex enterprise.6. To provide functional information about intrusions that do take place, allowing enhanced diagnosis, improvement, and correction of contributing factors

*C. Solution Overview*

Network Intrusion Detection Systems (NIDS) just analyze network traffic capture on the network section. NIDS may seek for either anomalous activity (anomaly based NIDS) or known intimidating patterns (signature based NIDS) on the network. Firewalls they do not normally block packets, but sensitive about the intrusion alert. This situation focuses on elusions over the signature of these systems. There are some troubles in network protocols that create state where endpoint systems process the packets generates a different demonstration of data in the NIDS and in the end system. If the design of the data in the NIDS and in the end systems are different then the evasion is successful. Sometimes it is not possible to detect the attacks. Thus, a possible appearance of new elusive techniques would be critical for systems that are supposed to be secure. This is the inspiration and motivation of the work, in which a new approach to watch for elusions over NIDS, giving a verification of concept showing how to perform deep packet inspection in NIDS using two publicly available datasets. This framework can be used for analyzing and Modeling and detecting the malicious behavior in the commercial NIDS. Computer security is defined as the protection of computing systems against threats to confidentiality, integrity, and availability [2]. The goal of Confidentiality (or secrecy) is that information is disclose only according to strategy, integrity means that information is not destroyed or despoiled and that the system performs properly availability means that system services are existing when they are needed.

Computing system refers as computer, computer networks and the information they handle security events come from different sources such as natural forces, accidents, failure of services (such as power) and people known as intruders. The categories of intruders are the external intruders who are unauthorized users of the system they attack, and internal intruders, who have permission to access the system with some limitations. The traditional prevention techniques such as user confirmation, information encryption, avoid programming errors and firewalls are used as the first line of guard for computer security. To manage the enormous amount of personal, public and critical data we always choose the Information Technology systems, which become a critical component in organizations Identifying such anomalous behavior and hostile actions for protecting those systems, which is one of the most important goal in security.

## III. PROPOSED WORK

The main aim is to develop a network based intrusion detection system based on modified Apriori approach for attack detection and test the input thus produced by the Apriori algorithm with the well known snort intrusion detection system, once a candidate sets for detecting different attacks are generated. These candidates in turn will be passed as inputs to the snort intrusion detection system for detecting different attacks.

In Figure 1 the proposed system flow is given where, the input to C4.5 algorithm using Weka tool. Weka tool is implementation of various classifying and clustering algorithm. C4.5 algorithm gives output as a tree. After that, adaboost algorithm is applied on output of C4.5. Adaboost algorithm contains steps like data labeling, training and testing. Data labeling will contain identification normal and attack packets. +1 means attack packet and -1 means normal packet. Training phase will contain initialization of parameters. Testing phase will contain real identification attack packets and classifying each detected attack under its category (Such as Dos attack, probe attack, U2R attack, R2Lattack). After that detection result and false alarm rate will get displayed.

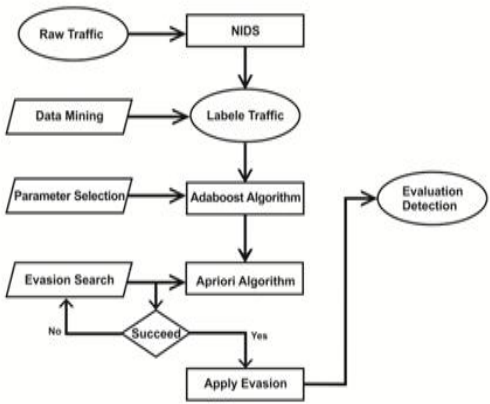

**Figure 1** Proposed System Structure

After this step modified apriori algorithm is used, which contain process of creation of rules for detecting attacks? After creating rules they are passed to snort. Snort is an open source IDS. Now this method will detect the packets in the network. It evades the packets by changing the rules. Detection output will get stored in text files. The workflow is depicted in the block diagram.

## IV. WORKFLOW

### A. KDD Cup DataSet

The KDD Cup 1999 dataset was derived from the 1998 DARPA Intrusion detection evaluation program prepared and managed by MIT Lincoln Laboratory. KDDCup99 Data set is used for Intrusion Detection and the development model is checked on the data set. The process of Artificial Intelligence for detection of intrusions is the method to build precise or correct IDS. To recognize misuse, anomaly detection and detecting key models are identified by using the rule based, Adaboost Algorithm and C4.5 algorithm techniques. The current data set does not have in the transaction format or does not have in the precise time information. We create the data records in the transaction format. Each transaction contains some records. In the Data set transactions are alienated by „###". The dataset in which an association rules is to be found is viewed as a set tuples, where each tuple consist a set of items. For example tuple {Dos, R2L, U2R} which comprises the three items, which are Smurf, R2L and U2R. Keeping the attack record register in mind, each item represents an attack happened in particular time period. Following are the different types of attacks:

1) *DOS attack*-It is the mechanism of making a computer resource unavailable, by blocking it from access intended for group of people. DOS attacks are generally targeted on high profile websites such as banks, payment gateways ping-of-death, teardrop, smurf, SYN flood.

2) *R2L attack*-It is attack performed by unauthorized person located outside the network with an attempt to hijack privileges of local users R2L, unauthorized access from a remote machine.

3) *U2R attack*-In this type of attack local host of a particular network tries to force fully hijack the privileges of super user like administrator. These types of attacks are very popular in UNIX systems unauthorized access to local superset privileges by a local unprivileged user.

4) *Probing attack*-It is a class of attack where an attacker scans a network to gather information or find known vulnerabilities.

### B. C4.5 Algorithm

C4.5 is an algorithm used to generate a decision tree developed by Ross Quinlan. C4.5 is an expansion of Quinlan's earlier ID3 algorithm. C4.5 builds decision trees from a set of training data, by the idea of information entropy. The training data is a set of already classified samples. Each sample is a vector where represent attributes or features of the sample. The training data is improved with a vector where signify the class to which each sample belongs. At each node of the tree C4.5 selects one attribute of the data that most successfully divide its set of samples into subsets enrich in one class or the further class. Its principle is the normalized information gain (difference in entropy) that results from chooses feature for divide the data. The attribute with the highest normalized information gain is chosen to make the decision. The C4.5 algorithm then recourses on the smallest sub lists. This algorithm has a little base case. All the samples in the list feel right to the identical class. When this happens, it just creates a leaf node for the decision tree saying to choose that class.

· None of the features provide any information gain. In this case, C4.5 gives a decision node upper the tree by way of the expected value of the class.

· Instance of previously-unseen class encounter. Again, C4.5 creates a decision node advanced the tree using the expected value.

## C. Adaboost Algorithm

AdaBoost means Adaptive Boosting. It is a machine learning algorithm developed by Yoav Freund and Robert Schapiro. It is a meta-algorithm and can be used in conjunction with many other learning algorithms to improve.

In some problems however it can be fewer susceptible to the over fitting problem than most learning algorithms. The classifiers it uses can be fragile (i.e., display a substantial error rate), but as extended their performance is not arbitrary (resulting in an error rate of 0.5 for binary categorization), they will improve the last model. Even classifiers with an error rate higher than would be predictable from a random classifier will be useful, since they will have pessimistic coefficients in the final linear combination of classifiers and hence behave like their inverses.

The reasons for using Adaboost algorithm are:
1) The AdaBoost algorithm is one of the most popular machine learning algorithms.
2) The AdaBoost algorithm corrects the misclassifications made by fragile classifiers, and it is less vulnerable to over fitting than most learning algorithms.
3) Data sets for intrusion detection are a heterogeneous mixture of categorical and continuous types.
4) If simple weak classifiers are used, the AdaBoost algorithm is very fast.

AdaBoost generates and calls a new weak classifier in each of a sequence of rounds. For each call a distribution of weights is updated that indicates the importance of examples in the data set for the categorization. On each round, the weights of each incorrectly classified example are rising and the weights of each correctly classified example are lower, so the new classifier focuses on the examples which have so far elude exact classification.

## D. Apriori Algorithm

One of the most popular data mining approaches is to find frequent item sets from a transaction dataset and derive association set of laws. A finding frequent item set (item sets with frequency larger than or equal to a user specified minimum support) is not trivial because of its combinatorial explosion. Once frequent item-sets are obtained, it is straightforward to create association rules with confidence larger than or equal to a user specified least confidence. Apriori is a seminal algorithm for finding frequent item-sets using candidate generation [1]. It is characterize as a level-wise total search algorithm using anti-monotonicity of item-group "if a nodes-group is not probable, any of its superset is by no way common". Through principle, Apriori presume that item-group is sort in lexicographic order. Assume that the group of frequent item-sets of size $k$ be $F_k$ and their candidates are $C_k$. Apriori initial scans the database and searches for frequent item-sets of size 1 by accumulating the count for each item and collecting those that satisfy the minimum support necessity. It then iterates on the following three steps and extracts all the frequent item-sets.

Pseudo code
$L1$ = {large 1-itemsets};
For($k = 2$; $L_{k-1} \neq \emptyset$; $k$++ ) do begin
$C_k$= apriori-gen($L_{k-1}$); //New candidates

their performance [6]. AdaBoost is adaptive in the sense that Subsequent classifiers built are tweaked in favor of those instances misclassified by earlier classifiers. AdaBoost is sensitive to noisy data and outliers.

For all transactions $t \in D$ do begin
$C_t$ = subset($C_k$, $t$);//Candidates contained in $t$
For all candidates $c \in C_t$
do $c$.count++;
End
$L_k$= { $c \in C_k$| $c$.count $\geq$ min-sup
}End
Return $L_k$;$\cup_k$

Apriori algorithm suffers from data complexity problems, i.e. for every step of candidate generation the algorithm has to scan the entire database, and as we are aware that the larger the database the difficult it is to scan completely, therefore candidate generation and candidate pruning are considered to be tedious as it involves bringing new data, after random unexpected intervals of time. We have to also take care that less memory should be utilized during the scanning process.

## V. RESULT

The aim of the proposed framework is to detect attacks on the network by using Apriori approach. By using this approach we will get the high detection rate and low false alarm rate. The two main objectives of the proof of concept presented are first to corroborate that Apriori Algorithm can be a good paradigm to model NIDS and second to find output is evasions over the NIDS. For that purpose, we have created a basic NIDS based on the C4.5 algorithm. Then the adaboost algorithm takes input as a decision tree from c4.5 algorithm. We had compared the detection rate of previous GP algorithm and apriori algorithm. The graph shows the comparison of detection rates in Figure 2.

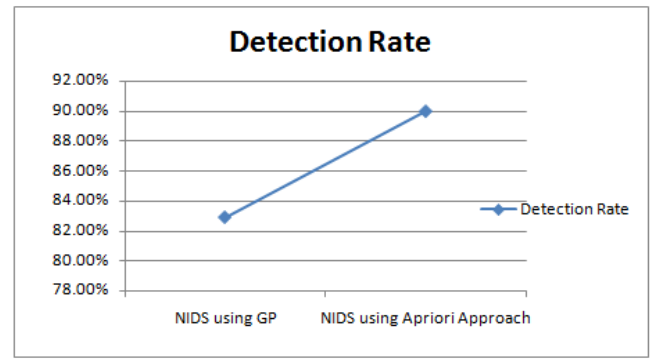

**Figure 2** Comparison of Detection Rate by using GP and Apriori Approach

The aim of paper is to find low false alarm rate. The comparison graph for the false alarm rates of NIDS by using GP and Apriori is as shown in Figure 3. The below graph shows that false alarm rate of NIDS by using apriori algorithm is very low as compare to NIDS by using GP.

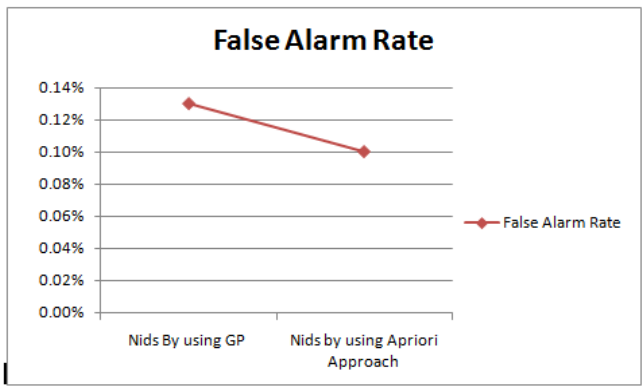

**Figure 3** Comparison of False Alarm rate by using GP and Apriori Approach

## VI. CONCLUSION AND FUTURE SCOPE

In the paper, we have implemented a new framework to look for evasions over a given NIDS. The framework results in high detection rate and low false alarm rate. This framework also gives the actual categorization of attacks on the network. For that purpose we are using the KDD 99 dataset and applying the signature Apriori algorithm which is well known and widely used for intrusion detection. This framework is used to detect the unknown attacks with high accuracy rate and high efficiency. This type of evade NIDS has very vast scope in future like one is to create our own dataset. The other is to analyze if this techniques can be applied straightly to model a commercial NIDS.